\documentclass[aps,prx,twocolumn,showpacs,superscriptaddress,longbibliography]{revtex4-1}  

\usepackage{braket}
\usepackage{soul}

\newcommand\underrel[2]{\mathrel{\mathop{#2}\limits_{#1}}}

\usepackage{graphicx}   
\usepackage{dcolumn}   
\usepackage{bm}            
\usepackage{amssymb} 
\usepackage{amsmath}  

\usepackage{color}
\usepackage{subfigure}
\usepackage{braket}
\usepackage{slashed}

\begin{document}

\title{Influencers identification in complex networks through reaction-diffusion dynamics}
\author{Flavio Iannelli}
\affiliation{
	Institute for Physics, Humboldt-University of Berlin, Newtonstra{\ss}e 15, 12489 Berlin, Germany
}
\email{iannelli.flavio@gmail.com}

\author{Manuel S. Mariani}
\affiliation{
	Institute of Fundamental and Frontier Sciences, University of Electronic Science and Technology of China, Chengdu 610054, PR China}
\affiliation{URPP Social Networks, Universit\"at Z\"urich, 8050 Z\"urich, Switzerland}
\affiliation{Department of Physics, University of Fribourg, 1700 Fribourg, Switzerland}

\author{Igor M. Sokolov}
\affiliation{
	Institute for Physics, Humboldt-University of Berlin, Newtonstra{\ss}e 15, 12489 Berlin, Germany
}

\date{\today}

\begin{abstract}
A pivotal idea in network science, marketing research and innovation diffusion theories is that a small group of nodes -- called influencers -- have the largest impact on social contagion and epidemic processes in networks. Despite the long-standing interest in the influencers identification problem in socio-economic and biological networks, there is not yet agreement on which is the best identification strategy. 
State-of-the-art strategies are typically based either on heuristic centrality measures or on analytic arguments that only hold for specific network topologies or peculiar dynamical regimes.
Here, we leverage the recently introduced random-walk effective distance -- a topological metric that estimates almost perfectly the arrival time of diffusive spreading processes on networks -- to introduce a new centrality metric which quantifies how close a node is to the other nodes. We show that the new centrality metric significantly outperforms state-of-the-art metrics in detecting the influencers for global contagion processes. Our findings reveal the essential role of the network effective distance for the influencers identification and lead us closer to the optimal solution of the problem.  
\end{abstract}

\maketitle

\section{Introduction}

Networks constitute the substrate for the spreading of agents as diverse as opinions~\cite{degroot1974reaching,friedkin1990social}, rumors~\cite{maki1973mathematical}, computer viruses~\cite{pastor2007evolution}, and deadly pathogens~\cite{pastor2015epidemic}.
Differently from classical epidemiological~\cite{hethcote2000mathematics} and collective behavior models~\cite{granovetter1978threshold}, which typically assume homogeneously mixed populations, the network approach assumes that agents can only spread through the links of an underlying network of contacts~\cite{pastor2015epidemic}.
Network-mediated spreading processes are ubiquitous: for example, online users transmit news and information to their contacts in online social platforms~\cite{bakshy2011everyone, pei2014searching, zhang2016dynamics}; individuals form their opinion and make decisions influenced by their contacts in social networks~\cite{degroot1974reaching,friedkin1990social,friedkin2017truth}; infected individuals can transmit infectious diseases to their sexual partners~\cite{eames2002modeling}.

A long-standing idea in network science, marketing research and innovation diffusion theories is that in a given network, a tiny set of nodes -- called \emph{influencers} -- have the largest impact on social contagion and epidemic spreading processes. 
Many studies have aimed to accurately identify~\cite{kempe2003maximizing,kiss2008identification,lu2016vital,Liu2016,refId0}, target~\cite{galeotti2009influencing,hinz2011seeding}, and assess the impact of~\cite{watts2007influentials,iyengar2011opinion} the influencers for marketing purposes.
Proper identification and targeting are vital for organizations to design effective marketing campaigns in order to maximize their chances of success~\cite{domingos2001mining, kempe2003maximizing,leskovec2007dynamics}, for policy-makers to design effective immunization strategies against infectious diseases~\cite{cohen2003efficient}, for social media companies to maximize the outreach of a given piece of information, such as a news or a meme~\cite{borge2012absence}.

\begin{figure*}
\centering
\includegraphics[width=0.8\textwidth]{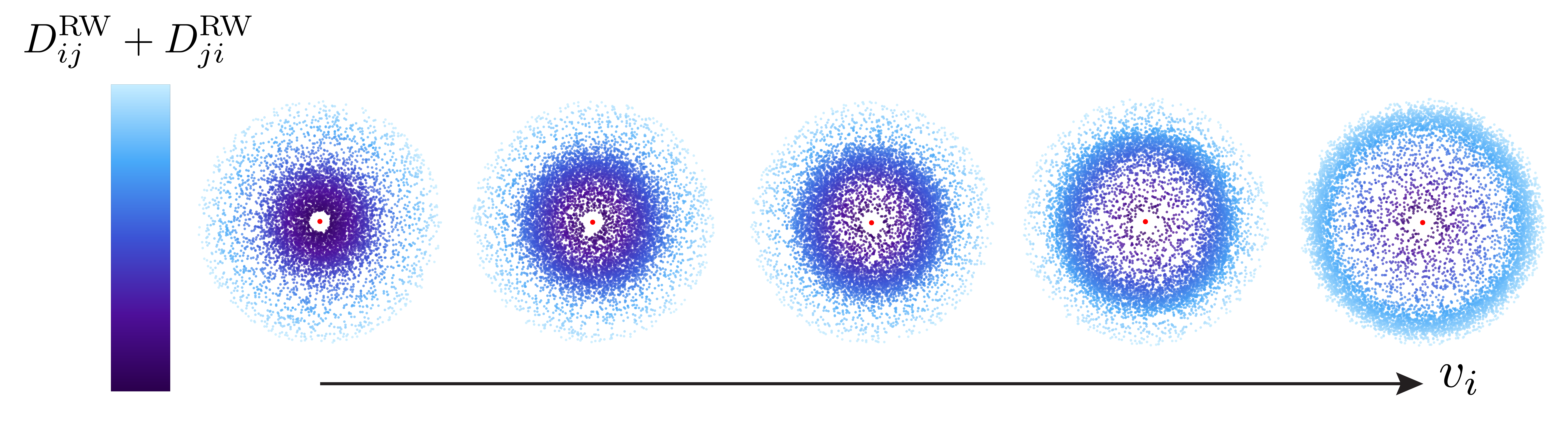}
	\caption{
 (Color online) Illustration of the ViralRank centrality in terms of the random-walk effective distance $D_{ij}^{\text{RW}}$ \cite{Iannelli} for different source nodes $\{i\}$ (central dots in the clouds). 
    The clouds of nodes around each seed node $i$ represent the other nodes $\{j\}$ in the network. Their radial distance from the center of the cloud is proportional to their total random-walk effective distance $(D_{ij}^{\text{RW}}+D_{ji}^{\text{RW}})$ with respect to the source node $i$; their color ranges from dark-blue (low distance) to white (high distance). The average effective distance yields the ViralRank score $v_i$ (horizontal axis). The cases depicted here evolve from a central node (small $v_i$, left side of the panel) which tends to be close to many other nodes, to a peripheral seed node (large $v_i$, right side of the panel) which tends to be far from the other nodes. 
 }
	\label{drawinga}
\end{figure*}

The influencers identification problem is typically studied by using epidemic spreading and social contagion models to simulate multiple independent realizations of spreading processes on real networks. Different processes are initiated by different "seed" nodes; the typical size of the outbreak generated by a given node quantifies its "ground-truth" spreading ability~\cite{kitsak2010identification,de2014role,bauer2012identifying,radicchi2016leveraging,lu2016vital}. One can thus compare different node ranking algorithms with respect to their ability to identify the nodes with the largest ground-truth spreading ability~\cite{kitsak2010identification,lu2016vital}. 
The seminal work by
Kitsak et al.~\cite{kitsak2010identification} showed that the nodes with the largest number of contacts ("hubs" in the network science literature~\cite{barabasi2016network}) are not necessarily the most influential spreaders, and nodes with fewer connections but located in strategic network positions can initiate larger spreading processes, see also discussion in \cite{PhysRevE.78.066109}.
Following Kitsak et al.~\cite{kitsak2010identification}, several network centrality measures~\cite{lu2016vital,liao2017ranking} -- originally aimed at quantifying individuals influence and prestige in social networks~\cite{katz1953new} -- have been compared with respect to their ability to identify the influential spreaders~\cite{chen2012identifying,borge2012absence,zeng2013ranking,liu2013ranking,de2014role,bauer2012identifying,lu2016h,radicchi2016leveraging, pastor2017topological,lawyer2015understanding}.
The results of this massive effort have been often contradictory, and there is not yet agreement on which is the best metric for the influencers identification.


The current lack of agreement on which metric best quantifies the spreading ability of the nodes can be ascribed to two main limitations of existing studies.
First, most of the proposed centrality measures do not consider the properties of the spreading dynamics in exam~\cite{bonacich1972factoring,kitsak2010identification,chen2012identifying,lu2016h}, or they are based on analytic arguments that are valid only for specific types of networks and spreading parameters~\cite{radicchi2016leveraging}.
As a result, the performance of these metrics strongly depends on network topology and on the parameters that rule the target epidemic process.
Second, existing works often restrict the comparison of the metrics performance to a limited number of parameter 
values~\cite{lu2016vital, radicchi2016leveraging}, which leaves it unclear how the relative performance of the metrics depends on model parameters.

In this article, we overcome both limitations. We introduce a new centrality metric, which we call \textit{ViralRank},
directly built on the random-walk effective distance for reaction-diffusion spreading processes~\cite{Iannelli}.
In particular, the ViralRank score of a node is defined as its average 
random-walk effective distance to and from all the other nodes in the network. The rationale behind this definition is that an influential spreader should be able to reach and to be reached quickly from the other nodes. As the random-walk effective distance quantifies almost perfectly the infection arrival time for any source and target node in reaction-diffusion processes~\cite{Iannelli}, we expect the average effective distance to accurately quantify how well a node can reach and be reached by the other nodes.

Our results show that ViralRank is the most effective metric in identifying the influential spreaders for global contagion processes -- both contact-network processes in the supercritical regime, and reaction-diffusion spreading processes. 
In contact networks, if the transmission probability is sufficiently large, ViralRank is systematically the best metric to quantify the spreading ability of a node.
We provide evidence that -- differently from what was previously stated~\cite{kitsak2010identification,radicchi2016leveraging} -- values of the transmission probability well above the critical point are relevant values to real spreading processes. 
In the metapopulation model, ViralRank is the best-performing metric for almost all the analyzed parameter values.
Besides, we show analytically that ViralRank can be written in terms of the classical Friedkin-Johnsen social influence model, introduced in~\cite{friedkin1990social} and recently used to predict individuals final opinions in controlled experiments~\cite{friedkin2016theory, friedkin2017truth}.
We also show that the Google PageRank~\cite{brin1998anatomy} score can be re-interpreted as the average of a specific \textit{partition function} built on the network effective distance.

Our findings demonstrate that the effective distance between pairs of nodes can be used to quantify the nodes
spreading ability significantly better than with existing metrics, bringing us closer to the optimal solution to the problem of identifying the influential spreaders for both contact-network and reaction-diffusion processes.

\begin{figure}
\centering
\includegraphics[width=0.4\textwidth]{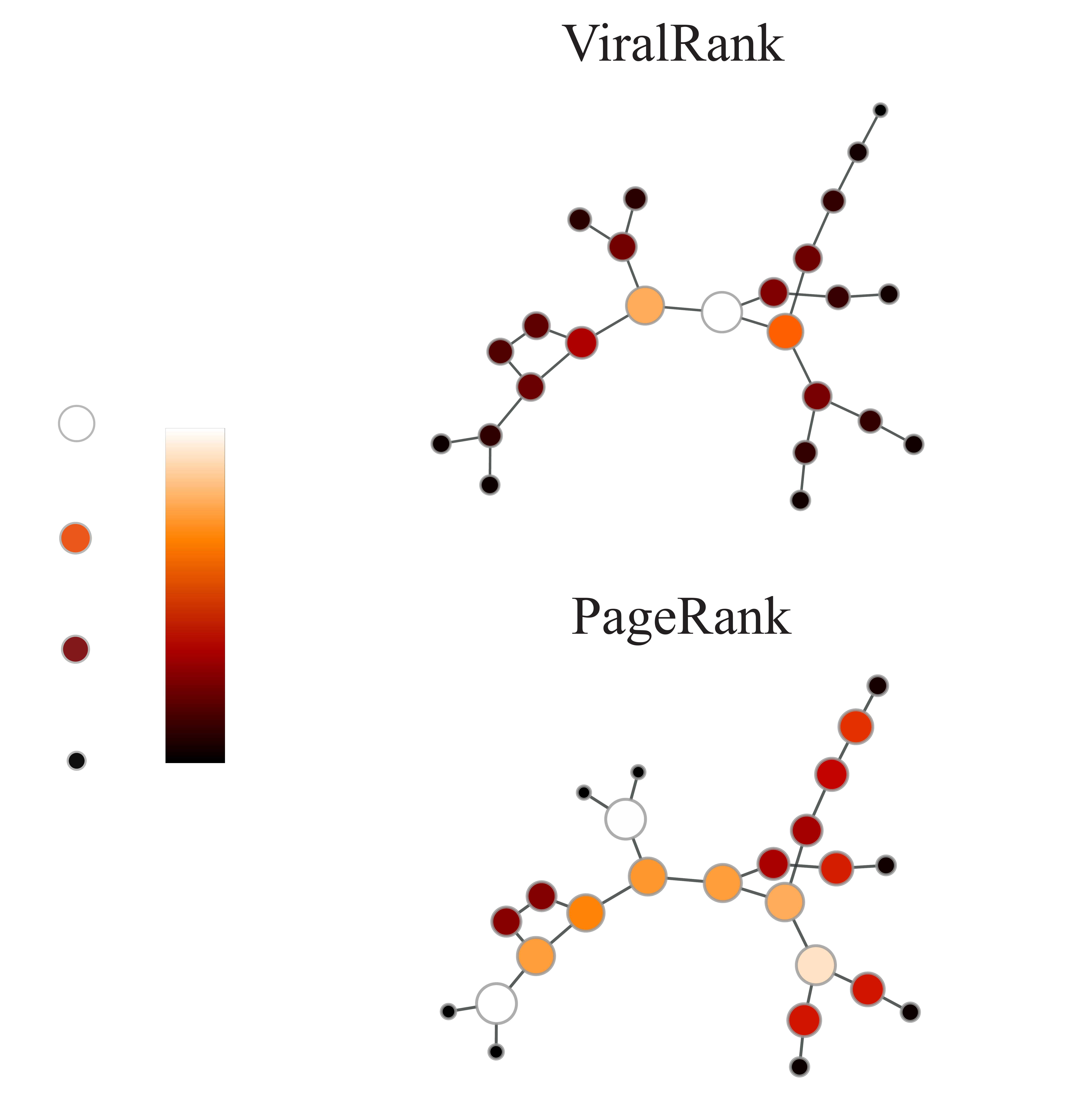}
	\caption{
(Color online) Comparison between ViralRank and PageRank with dumping parameter $0.85$ \cite{Gleich} for a toy small-world network with $N=25$ nodes. In line with~\cite{watts1998collective}, the network is built by starting from a ring topology where each node has $5$ neighbors and by rewiring each edge with probability $p=0.5$. The size of each node is proportional to the  value of the corresponding score normalized by the maximum score in the network, with color scale changing accordingly.
 }
	\label{drawingb}
\end{figure}




\section{Results}

We start by defining the new metric (ViralRank) and then validate it as a metric for the influential spreaders identification for contact-network and reaction-diffusion processes. \textit{Contact-network} models of spreading assume that individuals directly "infect" the individuals they are in contact with. Crucially, the topology of the underlying network of contacts plays a critical role in determining the size of the infected population~\cite{pastor2001epidemic, watts2002simple}. 
On the other hand, to describe global contagion processes, \textit{reaction-diffusion} models assume that individuals can infect the individuals that belong to the same population (reaction process) and in addition, infected individuals can move across adjacent locations (diffusion process). 


\subsection{ViralRank}

Previous works~\cite{brockmann2013hidden,Iannelli} have pointed
out that in order to predict the hitting time of a spreading process in geographically-embedded systems,  network topology and the corresponding weight flows play a  more fundamental
role than the geographical distance. 
The main idea behind ViralRank is to rank the nodes based on the random-walk effective distance $D_{ij}^{\text{RW}}(\lambda)$ between pairs of nodes which quantifies almost perfectly
the hitting time of a reaction-diffusion process on networks~\cite{Iannelli}.
Importantly, the calculation of $D_{ij}^{\text{RW}}(\lambda)$ only requires the network adjacency matrix $A_{ij}$ as input, whereas $\lambda$ is a parameter that depends on the spreading dynamics (see below).

We define the ViralRank score of a node $i$ as the average random-walk effective distance from all sources and to all target nodes in the network \footnote{We assume that the network is connected.} 
\begin{equation}
v_i(\lambda) = \frac{1}{N} \sum_{j} \left(D_{ij}^{\text{RW}}(\lambda) + D_{ji}^{\text{RW}}(\lambda)\right),
\label{viralrank}
\end{equation} 
where the effective distance is defined by~\cite{Iannelli} 
\begin{equation}
D_{ij}^{\text{RW}}(\lambda) = -\ln \Biggl( \sum_{k \ne j} \left( \mathbf{I}^{(j)}- e^{-\lambda} \mathbf{P}^{(j)}\right)^{-1}_{ik}  e^{-\lambda} p_{k}^{(j)}\Biggr)
\label{effdist}
\end{equation}
for $i\neq j$, whereas $D_{ii}^{\text{RW}}(\lambda)=0$.
The argument of the logarithm is a function that counts all the random-walks that start in $i$ and end when arriving in $j$ -- we refer to it as \textit{partition function}, see Appendix~\ref{app:interpretation}.
Here, $\mathbf{P}^{(j)}$ and $\mathbf{I}^{(j)}$ are the $(N-1)\times(N-1)$ submatrices of the Markov matrix \footnote{For weighted networks the weights have to be considered in place of $A_{ij}$.} $(\mathbf{P})_{ij} = A_{ij}/\sum_k A_{ik}$  and of the identity matrix $(\mathbf{I})_{ij} = \delta_{ij}$, respectively, obtained by excluding the $j$th row and $j$th column; $\mathbf{p}^{(j)}$ is the $j$th column of $\mathbf{P}$  with the $j$th component removed. 
The nodes are therefore ranked in order of \textit{increasing} ViralRank score: a node is central if it has, on average, small effective distance from and to the other nodes in the network \footnote{To compare ViralRank  performance with that of metrics that rank the nodes in order of \textit{decreasing} score (e.g., degree), we use $-v$. In this way, the nodes are again ranked in order of decreasing (yet increasing in modulus) score. To keep the terminology simple, we always refer to the correlation between $-v$ and  spreading ability as ViralRank performance.}.
As the nodes ranked high by ViralRank tend to have small effective distance from the other nodes, we expect them to generate larger epidemic outbreaks than peripheral nodes when they are chosen as the "seed" nodes of a spreading process (see Figure \ref{drawinga}).
Testing the validity of this hypothesis is one of the main goals of this paper.

For reaction-diffusion processes, the interpretation of $D_{ij}^{\text{RW}}(\lambda)$ as a proxy for the hitting time of the spreading agent makes the parameter $\lambda$ unambiguously determined by the transmission and recovery rates of the process  (see~\cite{Iannelli}).
For contact-network processes, a clear-cut criterion to choose $\lambda$ is lacking. Our analytic results (see Appendix~\ref{app:interpretation}) show that, in the limit $\lambda\to 0$ the ViralRank score of a given node $i$ reduces to the average mean first-passage time (MFPT) needed for a random walk starting in node $i$ to reach the other nodes, plus the MFPT   \footnote{This MFPT is also known as global MFPT~\cite{tejedor2009global}.} needed for a random walk starting in the nodes other than $i$ to reach node $i$. In the following, for contact-networks, we therefore consider the quantity $v_i = v_i(\lambda\rightarrow 0)$ as node $i$  ViralRank score. With this choice, a node $i$ is central if a random walk starting at node $i$ is able to quickly reach \emph{for the first time} the other nodes and, at the same time, it is well reachable from all other nodes.

In Appendices~\ref{app:fj} and ~\ref{app:pr}, we show that (1) there is a mathematical relation between ViralRank and the Friedkin-Johnsen (FJ) opinion formation model \cite{friedkin1990social}; (2) Google  PageRank can be also expressed, as ViralRank, in terms of a specific partition function. 
Our analytic computations reveal the two main differences between ViralRank and PageRank: (1) differently from the ViralRank score, the PageRank score does not depend logarithmically on its partition function, but linearly. This means that if a seed node $i$ is far from a node $j$ in the network, this will result in a small positive contribution to node $i$  PageRank score; by contrast, it will result in a large contribution (penalization) to its ViralRank score, proportional to $D_{ij}^{\text{RW}}$. (2) The specific partition function used by PageRank also includes the walks that hit several times the arrival nodes, which results in a poor estimate of the diffusion hitting time.

These two factors impair PageRank  ability to identify central nodes in networks. We show this by analyzing a toy Watts-Strogatz~\cite{watts1998collective} network with a clear distinction between central and peripheral nodes, see Fig. \ref{drawingb}.
The PageRank centrality~\cite{Gleich} gives a comparable score to peripheral nodes, located at the end of a branch, and central nodes, whereas ViralRank is able to clearly identify central nodes. In \cite{SM}, we show that PageRank is always outperformed by the degree centrality in the influential spreaders identification; for this reason, we do not show its performance here.

\subsection{Influential spreaders identification: Results for contact networks}

After having defined ViralRank and discussed its relation with PageRank and the FJ opinion formation model, we validate it as a metric for the influential spreaders identification. 
The metrics considered here for comparison are the following: degree centrality $k$, k-core centrality $k_c$~\cite{kitsak2010identification}, 
random-walk accessibility (RWA)~\cite{de2014role}, LocalRank (LR)~\cite{chen2012identifying} and the non-backtracking centrality (NBC)~\cite{martin2014localization}.
All these metrics are defined in Appendix~\ref{app:centralitymetrics}.

\paragraph*{Spreading dynamics.}

In this section, we consider \textit{contact-network} processes where the spreading agent is directly transmitted from an infected node to its susceptible neighbors.
More specifically, we consider a susceptible-infected-removed
(SIR) model, which is one of the most studied mathematical models for epidemic spreading \cite{pastor2015epidemic}. At each time step, each individual (node) can be in one of three states: susceptible, infected, or removed. 
Each infected node can infect each of its susceptible neighbors with probability $\beta$, and then infected nodes are removed from the dynamics with probability $\mu$. 
The process terminates when there is no infected node in the network and the disease cannot propagate anymore.
To assess the metrics performance we compare the scores they produce with the scores of the nodes by their spreading ability~\cite{kitsak2010identification,lu2016vital}. The spreading ability $q_i$ of node $i$ is defined as the average number of nodes in the removed state after the infection process has ended, given that the process was initiated by node $i$ -- i.e., node $i$ was the only infected node at time $t=0$.
For each node $i$, this average is based on $10^3$ independent realizations of the stochastic SIR dynamics described above.

For the SIR model, there exists a critical value (referred to as epidemic threshold \footnote{The epidemic threshold for the SIR model can be estimated within the degree-block approximation, i.e. assuming no degree correlations,  
as~\cite{dynamical} $\beta_c  = \braket{k}/(\braket{k^2}-\braket{k})$, where $k_i = \sum_j A_{ij}$ denotes the degree.}) $\beta=\beta_c$ such that the spreading process, once initiated, quickly dies out for $\beta<\beta_c$, whereas it infects a significant portion of the network, i.e. non-vanishing in the thermodynamic limit, for $\beta>\beta_c$. 
We expect the distance of $\beta$ from $\beta_c$ to significantly affect the relative metrics performance, an aspect that is typically not extensively investigated in existing works on the influential spreaders identification.
Below, we study how the metrics performance depends on $\beta/\beta_c$.

\paragraph*{Results.}

We first analyze synthetic networks composed of $N=100$ nodes and $L=189$ links. To uncover how network topology affects the metrics performance, we start from a network generated using the configuration  model \cite{bender1978asymptotic} with degree distribution following a power-law $\mathcal{P}(k) \sim k^{-\gamma}$, with exponent $\gamma = 2$, and we replace a fraction $p$ of its links with links that connect pairs of randomly selected nodes. In this way, we move continuously from a scale-free network ($p=0$) to a random (Poissonian) topology ($p=1$).

\begin{figure*}
	\centering
	{\includegraphics[width=0.8\textwidth]{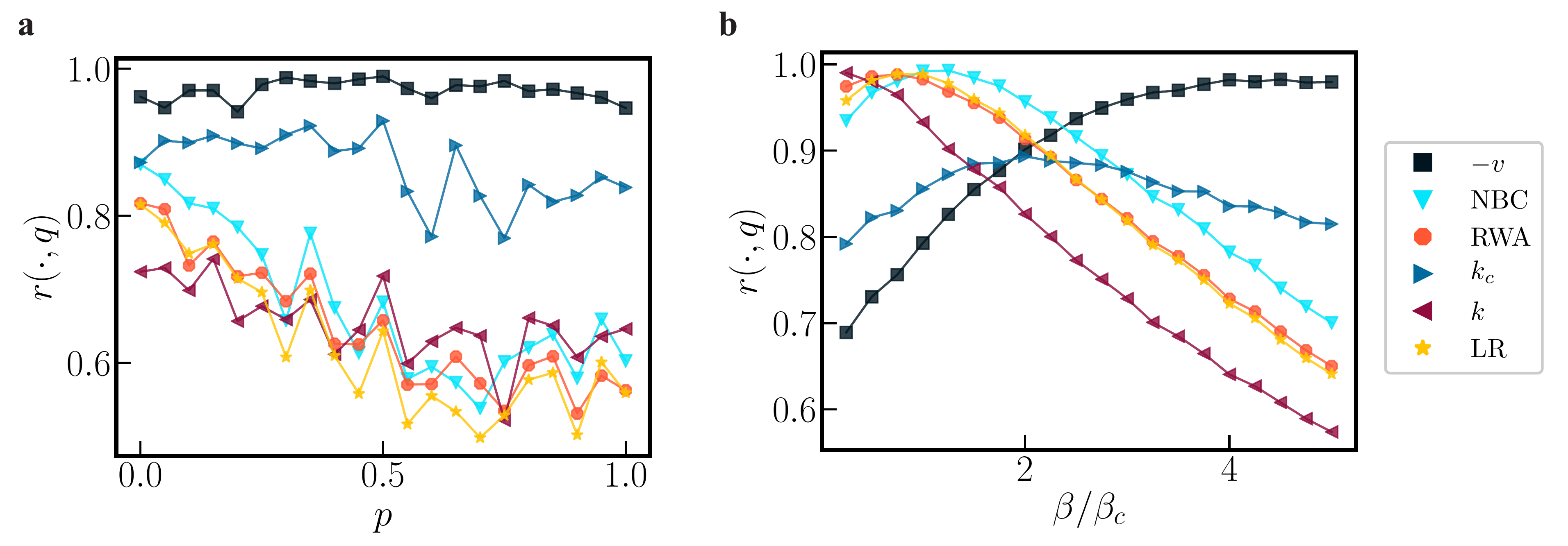}}
	\caption{(Color online) Correlation between nodes  centrality score and nodes  spreading ability $q$ in synthetic networks composed of $100$ nodes. (a) Pearson correlation coefficient between node centrality and $q$ as a function of the link rewiring probability $p$, at fixed $\beta/\beta_c=4$. The extreme points $p=0$ and $p=1$ correspond to a scale-free and to a Poissonian topology, respectively. (b) Pearson correlation coefficient between nodes  centrality and $q$ as a function of $\beta/\beta_c$, at fixed $p=0$ (scale-free topology). 
	}	\label{corr_synthetic}
\end{figure*}

Fig. \ref{corr_synthetic} (a) shows the Pearson correlation coefficient  $r(\cdot,q)$ between nodes  spreading ability $q$ and node score as a function of the shuffling probability $p$, for a fixed value of the ratio $\beta/\beta_c=4$ and for all the considered centralities. We find that all metrics besides ViralRank decrease their correlation with the spreading ability as the network topology becomes more homogeneous (i.e., as $p$ increases). This reflects the fact that for a random but homogeneous topology ($p=1$), the spreading ability spans a narrower range of values and, as a consequence, it becomes increasingly harder for the metrics to accurately estimate $q$. 
ViralRank is the best performing metric for all the $p$ values; nevertheless, we shall see in the following that the metrics  relative performance critically depends on $\beta$.

Fig. \ref{corr_synthetic} (b) shows the correlation $r(\cdot,q)$ as a function of $\beta/\beta_c$ for the scale-free network ($p=0$). 
First, we note that around the critical point $\beta = \beta_c$, LR, NBC and RWA all display a peak of maximum correlation with the spreading ability. This is in qualitative agreement with the fact that the NBC is expected to accurately estimate the size of the percolation giant component at the critical point \cite{radicchi2016leveraging}, for locally tree-like graphs; at the same time, it remains interesting that LR and RWA display a similar behavior.
This figure also shows that above the critical point $\beta_c$, there exists an \emph{upper-critical} value $\beta_u>\beta_c$ such that ViralRank is always the best performing metric for $\beta\ge\beta_u$. Real-data analysis shows that such point $\beta_u$ exists for all the analyzed empirical datasets (see below).

We note that there is a sensible decrease in the overall performance of all metrics as $\beta$ increases. This reflects the fact that as we approach the saturation value $\beta=1$, the distribution of nodes  spreading ability $q$ becomes narrower, making it harder  for the metrics to quantify $q$. Nevertheless, we emphasize that for values of $\beta$ as large as $\beta=7\,\beta_c$ in this synthetic network, we are still able to observe significant differences among the metrics  performance. This indicates that the influential spreaders identification in the super-critical regime is still a non-trivial problem, an aspect that will also emerge in real data.

To summarize, the results on synthetic networks show that in general the metrics relative performance critically depends on the heterogeneity of the underlying network topology and on the spreading parameters.
The previous results also suggest that ViralRank significantly benefits from the spreading process being super-critical.

We proceed by analyzing six empirical networks (see Table \ref{table} for a summary of their properties) in which we simulate the SIR spreading process: (a) 9/11 terrorists, (b) email, (c) jazz collaborations, (d) network scientists co-authorships,  (e) protein interactions and (f) Facebook friendships.
The meaning of the nodes and the links in the datasets and the datasets  properties are explained in Appendix \ref{app:datasets}. The results for six additional empirical datasets are shown in \cite{SM} and are in qualitative agreement with the results shown here.

\begin{figure*}
	\centering
	{\includegraphics[width=1.\textwidth]{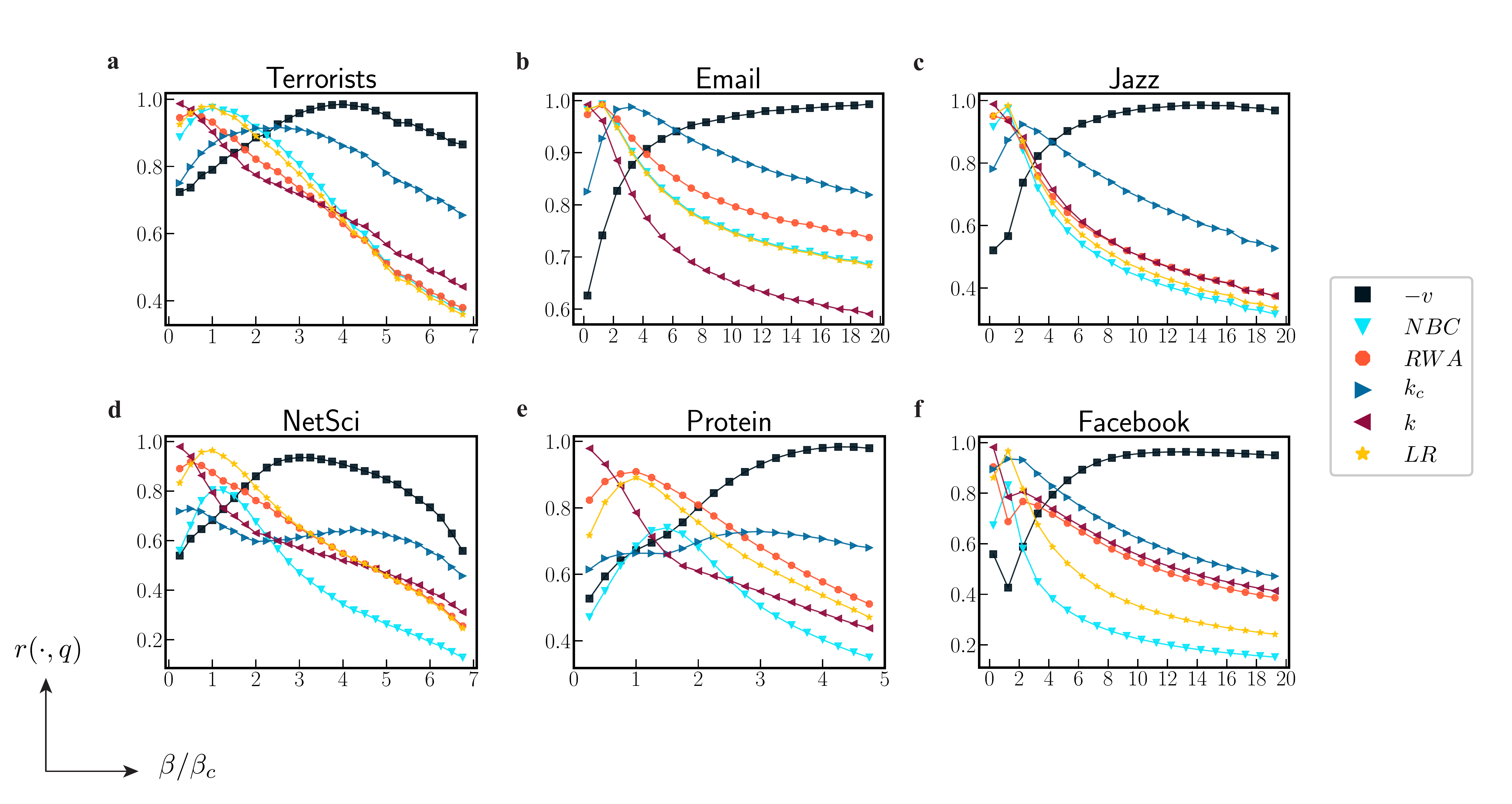}}
	\caption{(Color online) Correlation between nodes  centrality score and nodes  spreading ability $q$ in real networks.
		Pearson correlation coefficient between nodes  centrality and $q$ as a function of $\beta/\beta_c$ for the six datasets of Table~\ref{table}.
	}
	\label{corr_critical}
\end{figure*}

As in the case of synthetic networks, we find that for all the analyzed datasets, there exists a dataset-dependent value $\beta_u$ such that ViralRank is the best-performing metric for $\beta\ge \beta_u$, see Fig. \ref{corr_critical}. 
The value $\beta_u$ is always larger than $\beta_c$, which confirms that ViralRank is the most effective metric for the identification of influential spreaders for spreading processes in the supercritical regime. The largest ($\beta_u=6.5 \ \beta_c$) and smallest ($\beta_u=2 \ \beta_c$) values of $\beta_u$ are observed for email and network scientists co-authorships, respectively. 
By contrast, other metrics perform better in the vicinity of the critical point; which metric performs best in this parameter region critically depends on the considered dataset. At the critical point $\beta_c$, the best performing metrics are, for almost all datasets, the NBC and LR. Interestingly, for all the analyzed datasets, $k_c$ is the second-best performing metric (after ViralRank) in the supercritical regime.

These results demonstrate that among the existing metrics, there is no universally best-performing metric; the only consistent conclusion is that ViralRank outperforms all the other metrics for processes sufficiently far from criticality. Therefore, the optimal choice of a metric for identifying the influential spreaders critically depends not only on the considered dataset but also on the parameters of the particular spreading process that is chosen as ground truth.
Remarkably, in most of the analyzed datasets, not only ViralRank outperforms other metrics in the $\beta\ge \beta_u$ range, but it also approaches the perfect correlation with the spreading ability, $r(-v,q) \simeq 1$, for specific ranges of $\beta$ values within the supercritical region.
In the following we provide evidence that the favorable regime for ViralRank ($\beta\ge \beta_u$) is also the relevant one for real epidemic processes.

\begin{figure*}
	\centering
	\includegraphics[width=0.75\textwidth]{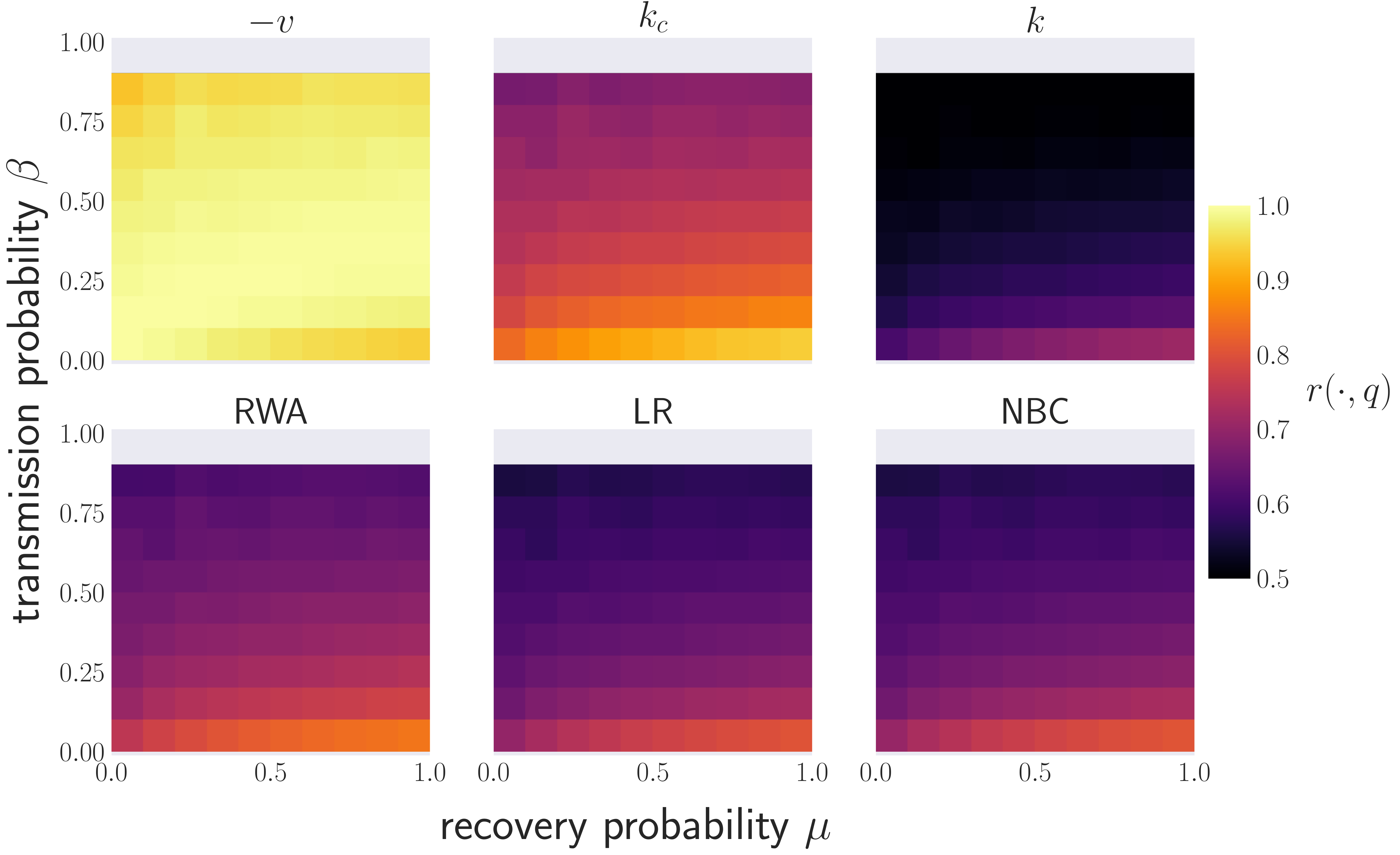}    
	\caption{\label{density}
		(Color online) Contact-network spreading model: a comparison between node centrality and node spreading ability $q$ in the whole parameter space, for the email network ($\beta_c = 0.0158\, \mu$) -- results for the other analyzed datasets are reported in \cite{SM}. The heatmap shows the Pearson  correlation coefficient $r(\cdot,q)$ between the nodes  centrality score and  spreading ability in the $(\beta,\mu)$ parameter space; the colors range from black ($r=0.5$) to yellow ($r=1$). 
	}
\end{figure*}

While ViralRank consistently outperforms the other metrics for $\beta\ge \beta_u$, we expect its performance to dwindle as $\beta$ approaches one. Indeed, for $\beta=1$, all the network nodes are eventually in the recovered state for any initiator of the process and, as a result, the nodes all have spreading ability equal to one.
To quantify the extent of the parameter region over which we are able to quantify the nodes  spreading ability, we study the complete parameter space $(\beta,\mu)$ of transmission and recovery probability. We find (Fig.~\ref{density} and \cite{SM}) that ViralRank is able to quantify the spreading ability, for a much larger parameter region than existing metrics. Remarkably, for the emails network (Fig. \ref{density}), the correlation between ViralRank and the spreading ability  $q$ is still larger than $0.95$ for values of $\beta$ as large as $\beta=0.9$ and still larger than $0.90$ even for $\beta = 0.99$. By contrast, for such large values of $\beta$, all the other metrics are essentially uncorrelated with $q$. Only at the saturation value  $\beta=1$ ViralRank loses its correlation with the spreading ability.

\begin{table}
	\begin{tabular}{l |r |r |r |r |r |r }
		Network & $N$ & $L$ & $D$ & $C$ & $\braket{k}$ & $\beta_u/\beta_c$
		\\ 
		\hline 
		\hline
		Terrorists & 62 & 152 & 5 & 0.49 & 4.90 & 2.50\\
		Email & 167 & 3250 & 5 & 0.59 & 38.92 & 6.50\\
		Jazz & 198 & 2742 & 6 & 0.62 & 27.70 & 4.25\\
		NetSci & 379 & 914 & 17 & 0.74 & 1.15 & 2.00 \\
		Protein & 1458 & 1948 & 19 & 0.07 & 2.08 & 2.25 \\
		Facebook & 4039 & 88234 &  8 & 0.61 & 43.69 & 4.75\\
		\hline 
	\end{tabular}
	\caption{Structural properties of the analyzed empirical networks: the different quantities represent the number of nodes ($N$) and links ($L$), the diameter ($D$), the clustering coefficient ($C$), the first ($\braket{k}$) moment of the degree distribution, and the upper-critical threshold ($\beta_u$) above which ViralRank outperforms all the other metrics, as a multiple of the SIR epidemic threshold $\beta_c$.
	}
	\label{table}
\end{table}

\paragraph*{Are real spreading processes above or below the critical point?}

The optimal performance of ViralRank for $\beta\ge \beta_u$ motivates the following question: how far are real spreading processes from criticality? To address this question, we use publicly available ranges $[R_0^{min}, R_0^{max}]$ of observed basic  reproductive numbers (see below), given in Table 10.2 of Ref. \cite{barabasi2016network} for a set of real diseases, and publicly available values of observed transmission rates for a set of computer viruses given in Table 2 of Ref. \cite{aron2002benefits}. We find that, by assuming the SIR dynamics on the analyzed datasets, not only real cases fall into the super-critical regime, but a number of them are in the region $\beta\ge \beta_u$ where ViralRank outperforms the other metrics in identifying influential spreaders.

\begin{figure}
	\centering
	\includegraphics[width=0.4\textwidth]{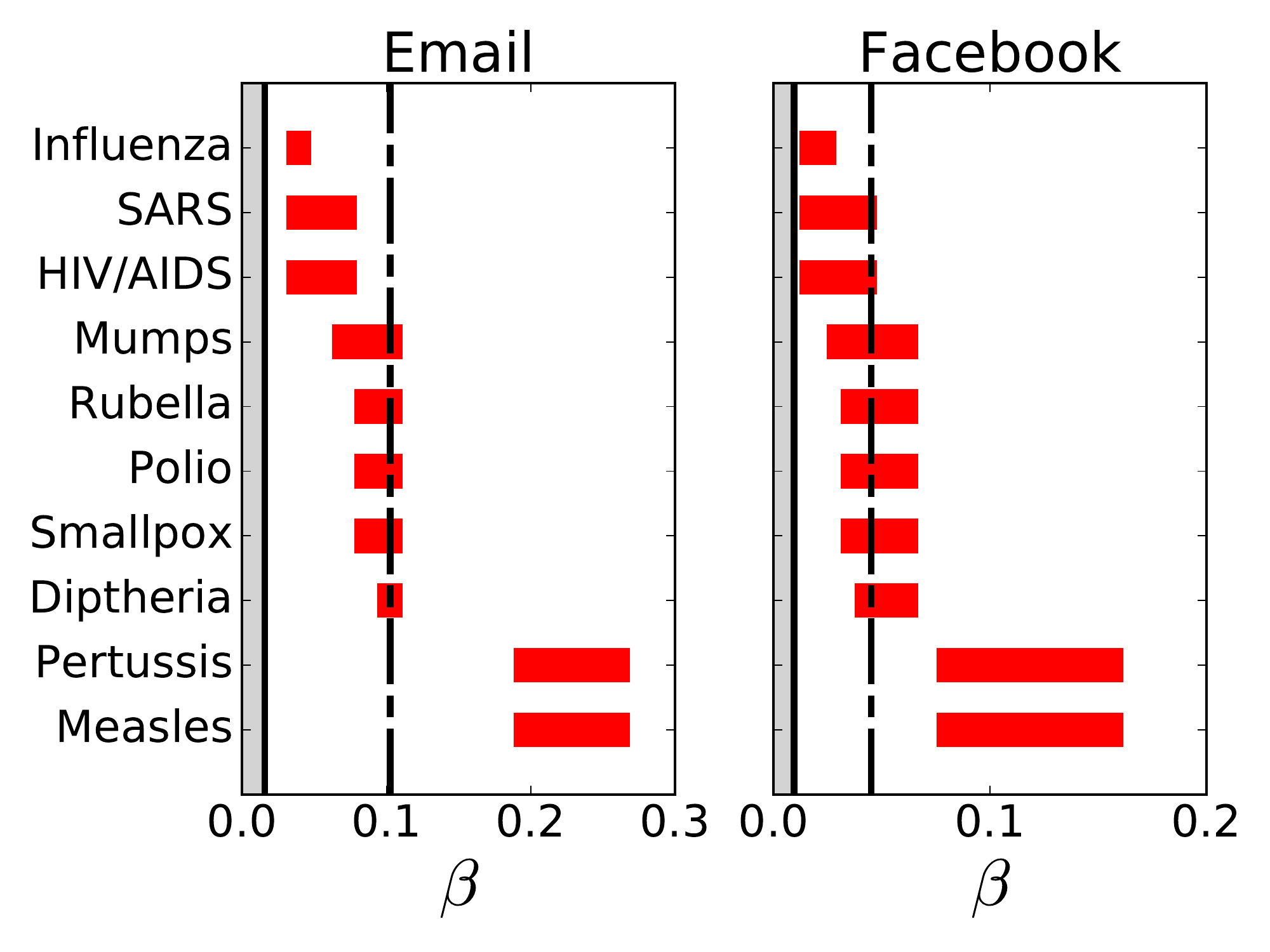}
	\caption{(Color online) Transmission-probability $\beta$ corresponding to real diseases for email and Facebook networks. The $\beta$ ranges (red horizontal bars) match the ranges $[R_{0}^{min}, R_{0}^{max}]$ observed for real diseases, taken from Table 10.2 of Ref. \cite{barabasi2016network}. By assuming $\mu=1$, the $R_0$ values are converted into $\beta$ values according to \cite{meyers2007contact} $\beta=R_0\,\braket{k}/(\braket{k^2}-\braket{k})$. The continuous and dashed vertical lines represent the epidemic threshold $\beta_c$ and the point $\beta_u$ such that ViralRank is the best-performing metric for $\beta\ge \beta_u$, respectively; gray and white colors fill the sub-critical and the super-critical interval, respectively.
	}    
	\label{fig:box}
\end{figure}

For a given disease, the basic reproductive number $R_0$ is defined as the number of secondary infections caused by a typical infected node in an entirely susceptible population \cite{anderson1992infectious}.
For the SIR model the heterogeneous mean-field approximation gives  \cite{meyers2007contact} $R_0\approx(\braket{k^2}/\braket{k}-1)\beta/\mu$, where $\braket{k}$ and $\braket{k^2}$ are the mean and variance of the degree distribution of the network of contacts. 
We can use this formula and the observed ranges $[R_0^{min}, R_0^{max}]$ to estimate, for each disease and each network of interest, the expected
lower and upper bounds (denoted as $\beta_{min}$ and $\beta_{max}$, respectively) for realistic values of $\beta$. 
We use this procedure to estimate the interval $[\beta_{min},\beta_{max}]$ for the ten diseases of Table 10.2 of Ref.  \cite{barabasi2016network} in two datasets, email and Facebook. The underlying assumption is that to some extent, these two networks can be considered as proxies for the social contacts that allow diseases to spread among individuals.
We find that for both datasets, real diseases fall in the super-critical regime, and often in the region $\beta\ge \beta_u$ where ViralRank outperforms the other metrics in identifying the influential spreaders, see Fig.~\ref{fig:box}.
For example, for the Facebook dataset, the minimum basic reproductive number (Influenza, SARS, HIV/AIDS, $R_0^{min}=2$) leads to $\beta_{min}=2\,\beta_c$, which lies still below $\beta_u=4.75 \,\beta_c$. On the other hand, the maximum value of $\beta$ for SARS and HIV/AIDS lies above $\beta_u$ ($\beta_{max}=5\,\beta_c$). 
The $\beta$ ranges for the diseases with the largest $R_0^{min}$ (Measles, Pertussis, $R_0^{min}=12$) lie well above $\beta_u$ ($\beta_{min}=12\,\beta_c$ and $\beta_{max}=17\,\beta_c$ for such diseases). 

Values of the transmission probability for some computer viruses \cite{kephart1993computers,kephart1997fighting,pastor2007evolution} can be found in Table 2 from \cite{aron2002benefits}. All the non-zero values reported in that table lie well above the critical point $\beta_c$ for the email dataset. The Word Macro virus ($\beta=0.7$) falls in the region where ViralRank significantly outperforms the other metrics; the Excel Macro virus ($\beta=0.1$) falls below but close to the point $\beta_u=0.103$, whereas the Generic.exe virus falls in the region where the $k$-core centrality is the best performing metric.

These examples indicate that by assuming a SIR dynamics, we expect the propagation of real diseases and computer viruses to be a super-critical diffusion process. We acknowledge that our argument above is simplified, as it assumes a free propagation of the disease (i.e., no external intervention aimed at limiting the impact of the disease) on an isolated population, which is unlikely to happen in the propagation of real diseases. Nevertheless, our assumptions are the same as those of all previous studies \cite{kitsak2010identification,lu2016vital,lu2016h,radicchi2016leveraging} that compared the performance of metrics for the influential spreaders identification using the SIR model.
Our argument therefore shows that, in the usual setting for benchmarking metrics for the influential spreaders identification 
borrowed from the epidemiology and network science literature~\cite{kitsak2010identification,lu2016vital},
the propagation of real diseases and computer viruses falls in the super-critical regime, and ViralRank is often the best performing metric in identifying the influential spreaders.
Besides, the supercritical region is also the most relevant from a marketing point of view: if the dynamics parameters force most of the spreading processes to die out quickly, it becomes virtually impossible for an influencer to initiate large-scale adoption cascades~\cite{watts2007influentials}.
A study of the problem in a more realistic setting goes beyond the scope of this work as it would require a more complex model of propagation, an accurate calibration of model parameters, and the possibility of external intervention (such as vaccination and travel restriction in the case of transportation networks).

To summarize, we have found that ViralRank systematically outperforms state-of-the-art 
centrality measures in the supercritical regime for contact-network spreading 
processes. In parallel, the poor performance at and below the critical point, shows the limitation of ViralRank. The decrease 
in performance can be easily explained in terms of the 
definition of network effective distances, upon which ViralRank is built. 
A basic assumption to define effective distances from a kinetic description of 
reaction-diffusion in interconnected subpopulations is that, information 
can reach \textit{all} nodes from any other node in a, possibly long but, finite time \cite{Iannelli}.
By extending this assumption to ViralRank, we average effective distances over 
all nodes, including those that are less likely to be infected for a subcritical 
 process that terminates after few time steps. 
In fact, for subcritical spreading processes the vast majority of nodes have 
practically zero probability to be reached by the infection, 
and in this case the average over all nodes in the definition of ViralRank 
is certainly not optimal.

\subsection{Influential spreaders identification: Results for metapopulation networks}

\paragraph*{Reaction-diffusion dynamics.}

While contact-network spreading processes can model the spreading of an infection within a network of individuals, in order to properly model global contagion processes, we need to take into account that multiple individuals, of different epidemiological compartments, can only interact with individuals that are located in the same geographical location.
This realization has motivated the study of \textit{metapopulation models} \cite{colizza07,colizza08}, where each node represents a geographical location that is occupied by a subpopulation composed of a subset of the metapopulation individuals. At each time step individuals can (1) interact with individuals located at the same node (\textit{reaction}); (2) travel across locations (\textit{diffusion}). Reaction-diffusion models of spreading are increasingly used to forecast the properties of epidemic outbreaks~\cite{balcan2010modeling,bajardi2011human,van2011gleamviz}, and to design and understand the systemic impact of disease containment strategies~\cite{bajardi2011human,tizzoni2012real}.

In the following, in line with previous studies~\cite{brockmann2013hidden,Iannelli}, we assume that the reaction dynamics is ruled by the fully-mixed SIR model; the generalization to arbitrary compartment models is obviously possible, but the SIR model often provides the sufficient level complexity necessary to describe real epidemic processes~\cite{Colizza2007}.

To simulate an epidemic, we use the weighted and undirected network of the $500$ most active commercial airports in the United States~\cite{colizza07}. A pair of airports is connected if at least one flight was scheduled between them in 2002; each link is weighted by the total number of passengers who flew between those two airports.
We assign to each node $j$ (airport) a subpopulation; airports are then connected to each others via the weighted adjacency matrix $W_{ij}$ that represents the undirected (averaged in both directions) flux of passengers between airports $i$ and $j$. 

The reaction-diffusion dynamics can be conveniently written for each compartment density $\rho_i^X$, where the place-holder variable $X$ can represent each of the three possible compartments: ${X}\in\{{S},{I},{R}\}$. The quantity $\rho^{I}_i(t)$ then can be viewed as the probability that node $i$ is infected at time $t$. 
The time evolution of the occupation densities consists of a sum of a diffusion term $\Omega(\{\rho^{X}_i\})$, known as the transport operator  \cite{dynamical}, and a reaction term $K^{X}(\beta,
\mu,\{\rho_i\})$ given by the fully mixed SIR model, which depends on the transmission and recovery rates $\beta$ and 
$\mu$. The ratio $R_0=\beta/\mu$ defines the basic reproductive number that serves as the control parameter of the system. 
Hence, we have a set of differential equations of the form  $\partial_t \rho_i^{X}  = \Omega(\{\rho^{X}_i\}) + K^{X}(\beta,\mu,\{\rho_i\})$.

To write the diffusion in terms of the compartmental densities only, i.e. without requiring the information about the subpopulation sizes, we make the following assumption. We assume the node strengths $s_i = \sum_{k} W_{ik}$ and the subpopulation sizes $\mathcal{N}_i$ as proportional via a constant \textit{diffusion rate} $\alpha =  s_i/\mathcal{N}_i$, which we set in our simulations to the fixed value $\alpha = 0.003 \ \text{d}^{-1}$, in units of days. The latter is also known in the literature as global mobility rate since it gives the fraction of moving agents per time step in the metapopulation \cite{brockmann2013hidden,Iannelli}. 
With the above assumption the transport operator can be written without the explicit dependence on the subpopulation size as $\Omega(\{\rho^{X}_i\}) = \alpha \sum_{k} P_{ik} \left(\rho^{X}_k  - \rho^{X}_i \right)$, where $P_{ik} = W_{ik}/\sum_j W_{ij}$ is the transition probability matrix.
The strength vector is then the equilibrium distribution of the Markov chain with states defined by the nodes.

The full metapopulation SIR dynamics then reads
\begin{eqnarray}
\begin{cases}
{\partial_t \rho_i^{S}} = \Omega(\{\rho^{S}_i\}) - \beta  \rho^{S}_i 
\rho^{I}_i \\
\\
{\partial_t \rho_i^{I}} = \Omega(\{\rho^{I}_i\}) + \beta  \rho^{S}_i 
\rho^{I}_i - \mu  \rho^{I}_i
\end{cases}
\label{RD}
\end{eqnarray}

\paragraph*{Identification of influential subpopulations.}

Despite the growing interest in reaction-diffusion processes~\cite{colizza2006role, Colizza2007,balcan2010modeling,brockmann2013hidden}, also spurred by their application to disease forecasting~\cite{van2011gleamviz}, the identification of influential spreaders for such dynamics has attracted less attention compared to the analogous problem for  contact networks.
Here, we fill this gap by comparing different centrality measures with respect to their ability to identify those airports that are able to infect a large portion of the network in a relatively short time.
To simulate the epidemic, we numerically integrate the set of  non-linear differential equations \eqref{RD}.

Importantly,  non-trivial dynamics in this model is obtained only above the epidemic threshold $R_0=\beta/\mu=1$, where all nodes will eventually contain at least one infected individual after a sufficiently long time. This, however, makes it impossible to quantify the nodes  ground-truth spreading ability by measuring the asymptotic number of nodes with at least one infected individual. 
To avoid this, we halt the simulations at a given threshold time $t_{max}(R_0)$.
The threshold time $t_{max}$ is set to half of the characteristic  time for travel, which is estimated by the inverse of the diffusion rate $\alpha$ and, since higher transmission rates correspond to lower infection hitting times, normalized by the basic reproductive number of the infection; i.e. $t_{max}(R_0) = (2R_0 \alpha)^{-1}$.
The results presented here are little sensitive to the exact choice of $t_{max}$ as long as $t_{max}$ is sufficiently large \cite{SM}.
The ground truth spreading ability is the fraction of subpopulations $\omega_i(t_{max})$ that contain at least one infected individual at time $t_{max}$, given that $i$ initiated the process. 
The performance of a metric is then quantified by the correlation between the scores it produces and the epidemic prevalence $\omega_i(t_{max})$.

\begin{figure*}
\centering
   {\includegraphics[width=0.8\textwidth]{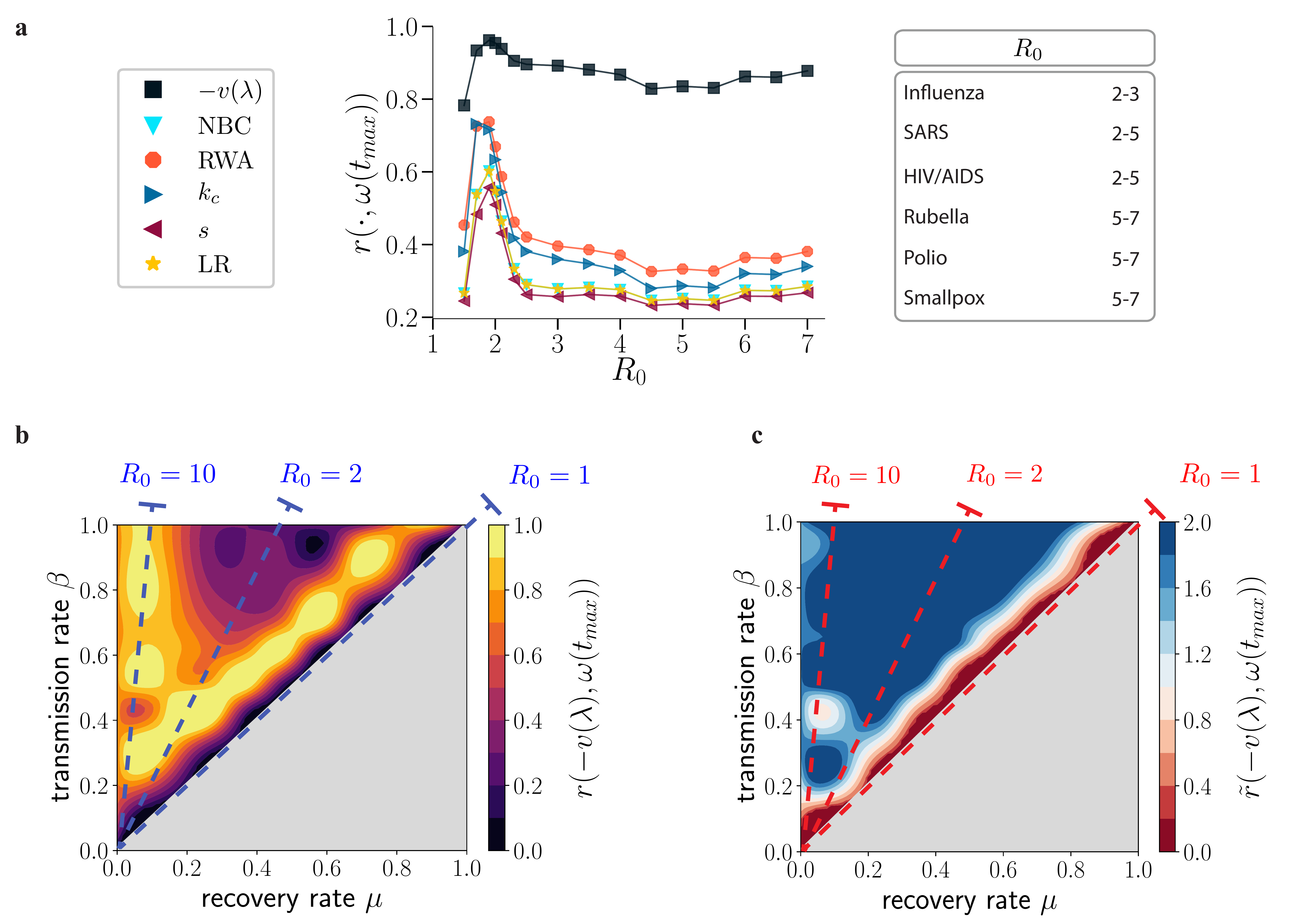}}
    \caption{(Color online) Metapopulation spreading model: A comparison between nodes centrality and prevalence $\omega(t_{max})$ for the U.S. domestic flights. (a) Pearson  linear correlation between nodes  centrality and $\omega(t_{max})$ as a function of the basic reproductive number, at fixed recovery rate $\mu=0.2 \ \text{d}^{-1}$, in unit of days. The inset shows the known $R_0$ values for some real diseases (from Table 10.2 in~\cite{barabasi2016network}). (b) Pearson  linear correlation $r(-v(\lambda),\omega(t_{max}))$ between ViralRank score and epidemic prevalence in the  non-trivial section of the parameter space $\{\beta>\mu\}$. (c) Ratio $\widetilde{r}$ between the correlations with prevalence of the score produced by ViralRank and the score obtained by the best performing metric (RWA), ViralRank excluded. The dashed lines in panels (b-c) mark the lines of constant basic reproductive number.
 }
 \label{meta}
\end{figure*}

\begin{figure}[]
	\centering
	{\includegraphics[width=0.45\textwidth]{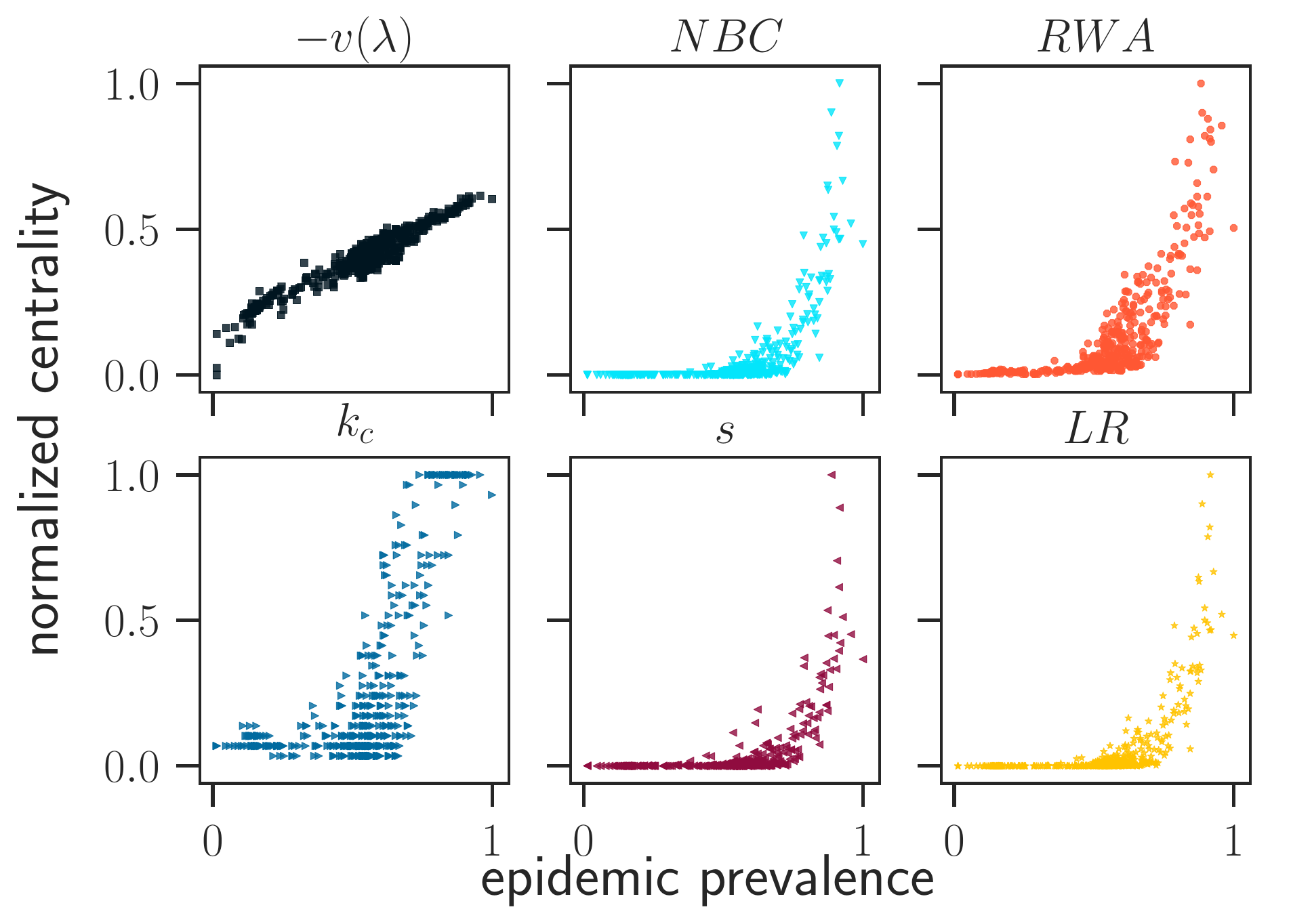}}
	\caption{(Color online) Scatter plot of the nodes  centrality scores  as a function of the prevalence $\omega(t_{tmax})$ at time $t_{max} = (2\alpha R_0)^{-1}$ for $R_0 = 2.0$ and $\alpha  = 0.003 \ \text{d}^{-1}$. For each axis, the values are normalized by the corresponding maximum value. 
	}
	\label{scattermeta}
\end{figure}

\paragraph*{Results.}

The definition of ViralRank for contact networks takes into account a formal limit of vanishing $\lambda$. 
In this limit, the ViralRank score of a node is equal to the average mean first-passage time from and to the other nodes. By contrast, for  metapopulations with the SIR reaction scheme, the parameter $\lambda$ has a direct relation with the dynamics parameters \cite{Iannelli}
\begin{equation}
\lambda(R_0,\mu,\alpha) = \ln\left[  \frac{(R_0-1)\mu}{\alpha} e^{-\gamma_e}\right], 
\label{lambdaopt}
\end{equation}
where $\gamma_e$ the Euler-Mascheroni constant. 
This value guarantees that the effective distance $D_{ij}^{\text{RW}}(\lambda)$ is highly correlated with the hitting time of the SIR reaction-diffusion process; as a consequence, for $\lambda=\lambda(R_0,\mu,\alpha)$, ViralRank is an accurate proxy for the average hitting time in the metapopulation. 

Inverting equation \eqref{lambdaopt} yields $R_0 = 1+ {\alpha}/{\mu} e^{\lambda + \gamma_e}$. 
Thus, in order to have a positive $\lambda$, condition necessary for the random-walk effective distance to be well defined,  we additionally require that the basic reproductive number in our simulations always satisfies $R_0 \ge 1+ {\alpha}/{\mu} e^{\gamma_e}$. However, this additional constraint only excludes a tiny interval of values from our analysis; for example, when $\mu=0.2 \ \text{d}^{-1}$ the threshold is given by $R_0 \ge 1.027$.

We compare the performance of all the previously considered centrality measures, by replacing the degree centrality with the strength $s_i=\sum_j W_{ij}$. 
We find that the ViralRank centrality $v_i(\lambda)$, with $\lambda$ given by equation \eqref{lambdaopt},  outperforms all the other metrics for almost all the values of $R_0$ by a great margin.
The correlation between the scores by the centrality measures and the prevalence $\omega(t_{max})$ as a function of the basic reproductive number $R_0$ (with fixed $\mu=0.2 \ \text{d}^{-1}$, in unit of days) is shown in Fig. \ref{meta} (a).
ViralRank is by far the best-performing metric for all the analyzed $R_0$ values. The scatter plots between the centrality scores and epidemic prevalence $\omega(t_{max})$ normalized by the respective maximum scores are reported for $R_0=2$ in Fig. \ref{scattermeta}, with ViralRank approaching the correlation $r(-v(\lambda),\omega(t_{max}))=0.95$.
The second-best performing metric is RWA, followed by $k_c$.

The observed performance advantage of ViralRank can be ascribed to the fact that differently from the other metrics, ViralRank built directly on an accurate estimate of the hitting time for reaction-diffusion processes on  networks~\cite{Iannelli}.
By extending the analysis to the whole non-trivial region $\beta>\mu$, the correlation between ViralRank and the epidemic prevalence stays larger than $r=0.8$ for a large portion of the accessible space (Fig. \ref{meta} (b)), and ViralRank is by far the best-performing metric in the whole probed space (Fig.~\ref{meta} (c)), apart from a confined region close to the diagonal $\beta=\mu$.
Importantly, as all real diseases reported in Table 10.2 of Ref.~\cite{barabasi2016network} have $R_0\geq 2$, they all fall into the parameter region where ViralRank significantly outperforms all the other metrics -- the region above the dashed line $R_0=2$ in Fig. \ref{meta} (b) and Fig. \ref{meta} (c).

\section{Discussion}

In this work, we have introduced a new network centrality, called ViralRank, which quantifies the spreading ability of single nodes significantly better than existing state-of-the-art metrics
for both contact-networks and reaction-diffusion spreading in the supercritical regime.
Our work is the first one that builds a centrality measure on analytic estimates of random-walk hitting times~\cite{Iannelli} and, at the same time, extensively validates the resulting centrality as a method to identify influential spreaders.
We make the code to compute ViralRank available at
\url{https://github.com/kunda00/viralrank_centrality}.
Besides, we have connected ViralRank to the well-known Friedkin-Johnsen opinion formation model~\cite{friedkin1990social}, and pointed out its difference with respect to the popular PageRank algorithm \cite{Gleich}.

Differently from most existing studies, our analysis involved the study of the whole parameter space of the target spreading dynamics. 
Our work emphasizes that differently from the common belief, the problem of identifying the influential spreaders in the supercritical regime is important for two main reasons. First, differently from what was previously thought~\cite{kitsak2010identification,radicchi2016leveraging}, there are large differences between the metrics  performance in this regime that are revealed by our analysis. Second, and most importantly, if we assume the SIR dynamics, the propagation of
real diseases and computer viruses falls always in the supercritical parameter region. 
This points out that while studying the spreading at the critical point remains an important theoretical challenge~\cite{radicchi2016leveraging}, supercritical spreading processes are in fact likely to be of practical relevance for applications to real spreading processes.

We conclude by outlining future research directions opened by our methodology and results. 
It remains open to extending the effective distance~\cite{Iannelli} and ViralRank to temporal networks. This might be of extreme practical relevance inasmuch as real networks exhibit strong non-markovian effects which in turn heavily impact the properties of network diffusion processes~\cite{rosvall2014memory,Scholtes,holme2016temporal}.
Besides, the SIR model provides a realistic yet simplified model of real diseases  spreading. Extending our results to more realistic spreading models is an important direction for future research; to this end, it will be critical to calibrate the spreading simulations with the parameters observed in real epidemics. 
While our work focused on a widely-used epidemic spreading model, an extensive validation of the metrics for social contagion processes~\cite{maki1973mathematical,borge2012absence} remains elusive, yet important direction for future research.

Our paper focused on the identification of \textit{individual} influential spreaders, in the sense that the simulated outbreaks always started from a single seed node.
Identifying a set of multiple influential spreaders might require different methods with respect to those used to identify individual influential spreaders~\cite{morone2015influence, lu2016vital}. Extending our results to spreading processes simultaneously initiated by more than one node is a non-trivial problem for future studies, yet relevant for real-world applications (such as targeted advertising and disease immunization) where it is typically more convenient to target a large number of potential influencers~\cite{bakshy2011everyone}.

Finally, ViralRank leads us closer to the optimal solution of the influential spreaders identification in the supercritical regime. While our results suggest that this regime is relevant for real spreading processes, it remains open to design, if at all possible, a universally best-performing metric that provides an optimal identification performance both in the supercritical and in the critical regime. For the SIR  model, our findings confirm that the non-backtracking centrality~\cite{radicchi2016leveraging} and LocalRank~\cite{lu2016vital} are highly competitive around the critical point, yet their performance declines quickly in the supercritical regime. By contrast, the $k$-core centrality provides a better performance -- yet sub-optimal with respect to ViralRank -- in the supercritical regime. Understanding whether the effective distance can be used to build a centrality metric that is also competitive around the critical point is an intriguing challenge for future studies.

\appendix

\section{ViralRank: interpretation and small $\lambda$ expansion}
\label{app:interpretation}

Let us write the random-walk effective distance \eqref{effdist} as  
\begin{equation}
D^{\text{RW}}_{ij}(\lambda) = -\ln Z_{ij}(\lambda),
\label{cm}
\end{equation}
where 
\begin{equation}
Z_{ij}(\lambda)=\sum_{n=1}^{\infty} e^{\ln H_{ij}(n)} e^{-\lambda n} = \braket{e^{-\lambda n_{ij}}},
\label{zeta}
\end{equation}
for $i \ne j$ and $Z_{ii}(\lambda)=1$.
In the last equation, $H_{ij}(n)$ is the hitting-time probability of a random walk with transition probability matrix $P_{ij} = A_{ij}/\sum_k A_{ik}$, obtained by normalizing the adjacency matrix $A_{ij}$, and $n_{ij}$ is the random-walk hitting time~\cite{norris1998markov}. The probability $H_{ij}(n)$ can be defined recursively as~\cite{norris1998markov} $H_{ij}(n) = \sum_{k\ne j} P_{ik} H_{kj}(n)$. The average $\braket{\dots}$ in  \eqref{zeta} is taken over all the random-walk realizations of length $n$ weighted by the probability $H_{ij}(n)$ that selects only those walks that terminate once $j$ is reached.

From equation \eqref{cm} an interesting analogy with thermodynamics emerges. The constant  $\lambda$ can indeed be interpreted as the inverse temperature. Correspondingly, $Z_{ij}(\lambda)$ is the  \textit{partition function}, and the effective distance corresponds to a reduced free energy per temperature.
Each walk-length $n$ in the partition function \eqref{zeta} is in one-to-one correspondence with a single internal energy level of the system; $H_{ij}(n)=e^{\ln H_{ij}(n)}$ quantifies the \textit{relative weight} of the configurations of energy $n$ -- i.e. the walks of length $n$ that terminate in $j$. 
Additionally, since $H_{ij}$ is a probability, the (microcanonical) entropy $\mathcal{S}^{\text{mic}}_{ij}(n) = \ln H_{ij}(n)$ of the energy level $n$ can be interpreted as the self-information \cite{cover2012elements} associated to the outcome of a random walker hitting node $j$ for the first time after $n$ steps starting from $i$.
The total internal energy is then given by the average of the hitting time dampened by a decreasing exponential $\mathcal{U}_{ij} = \braket{n_{ij} e^{-\lambda n_{ij}}}/\braket{e^{-\lambda n_{ij}}}$, with the partition function at the denominator. 
The canonical entropy is obtained as $\mathcal{S}_{ij} = \lambda\mathcal{U}_{ij}-\lambda\mathcal{F}_{ij}$, where $\lambda\mathcal{F}_{ij} =  D^{\text{RW}}_{ij}(\lambda) = -\ln \braket{ e^{-\lambda n_{ij}}}$ is the reduced free energy per temperature. 
Using the expression of the effective distance in terms of the cumulants $\braket{n_{ij}}_c^k$ of the hitting time~\cite{Iannelli} 
\begin{equation}
D^{\text{RW}}_{ij}(\lambda)=\sum_{k=1}^\infty (-1)^{k+1} \frac{\lambda^k\braket{n_{ij}}_c^k }{k!} ,
\label{expansion}
\end{equation}
the small-$\lambda$ expansion of node $i$  ViralRank score reads (up to a normalization constant)
\begin{align}
v_i 
\underrel{\lambda \rightarrow 0}{\approx} 
\lambda  \sum_j \left(\braket{n_{ij}} +\braket{n_{ji}} \right) 
+ \mathcal{O}(\lambda^2).
\label{hight}
\end{align}
Here $\braket{n_{ij}}$ is the mean-first passage time (MFPT) from $i$ to $j$ defined recursively as $\braket{n_{ij}} = 1 + \sum_{k\ne j} P_{ik} \braket{n_{kj}}$ if $i\ne j$, zero otherwise~\cite{kemeny1960finite}. 

In light of the analogy with thermodynamics outlined in the previous paragraph, as $\lambda$ can be interpreted as an inverse temperature, the ViralRank expression \eqref{hight} can be interpreted as a high-temperature expansion~\cite{parisi1988statistical}. In this limit, the internal energy reduces to the MFPT, whereas the higher-order terms in the expansion \eqref{expansion} give a vanishing contribution.
The small-$\lambda$ expansion shows that in the limit $\lambda\to 0$, apart from a uniform factor $\lambda$, node $i$  ViralRank score tends to the average MFPT from and to the rest of the network 
\begin{align}
\widetilde{v}_i \, {\approx} \,
 \sum_j \left(\braket{n_{ij}} +\braket{n_{ji}} \right).
\end{align}

\section{The relation between the Friedkin-Johnsen (FJ) opinion formation model and ViralRank}
\label{app:fj}

In the Friedkin-Johnsen (FJ) linear model of opinion formation in networks \cite{friedkin1991theoretical}, each node $i$ starts with an opinion $f_i$, with $\sum_i f_i=1$, and recursively updates it according to the linear iterative equation
\begin{equation}
\mathbf{y}(t+1)=c\,\mathbf{U}\,\mathbf{y}(t)+(1-c)\,\mathbf{f},
\label{FJeq}
\end{equation}
where $c$ is a model parameter, and $\mathbf{U}$ denotes a row-stochastic interpersonal influence matrix.
The final opinion $y_i$ of a node $i$ is linearly determined by the initial opinions $f_j$ of all the other nodes $\{j\}$ through the linear relation $\mathbf{y}(c|\mathbf{f})=\mathbf{V}\,\mathbf{f}$, where $\mathbf{V}(c)=(1-c)(\mathbf{I}-c\, \mathbf{U})^{-1}$. The matrix $\mathbf{V}$ can therefore be interpreted as the total interpersonal effects matrix~\cite{friedkin1991theoretical}.

In the following, we set $\mathbf{U=P}$, i.e. we assume that the interpersonal influence is completely determined by the network transition matrix $P_{ij} = A_{ij}/\sum_kA_{ik}$, where $A_{ij}$ is the adjacency matrix. 
Families of centrality measures can be constructed from the matrix $\mathbf{V}$. An important one, referred to as total effects centrality by Friedkin~\cite{friedkin1991theoretical}, defines node $j$  score as $\pi_j={N}^{-1}\sum_i V_{ij}$. As $V_{ij}$ represents the total interpersonal influence of $j$ on $i$, $\pi_j$ represents the average effect of node $j$ on the other nodes. 
Interestingly, as pointed out by Friedkin and Johnsen~\cite{friedkin2014two}, in the case of interest here ($\mathbf{U}=\mathbf{P}$), this metric is exactly equivalent to Google  PageRank~\cite{brin1998anatomy}.

Component by component, the FJ model \eqref{FJeq} reads
\begin{equation}
y_i(t+1)= \frac{c}{k_i}\,\sum_m A_{im}\,y_m(t)+(1-c)\,f_i.
\label{fj}
\end{equation}
The previous equation has a simple interpretation: each node starts with an opinion $f_i$, and recursively updates it by averaging its neighbors  opinions.
To connect the FJ model with ViralRank, it is instrumental to consider a $(N-1)\times (N-1)$ reduced matrix $\textbf{P}^{(j)}$ obtained from $\textbf{P}$ by removing the $j$-th row and column. The FJ opinion formation process associated with the reduced matrix $\textbf{P}^{(j)}$ reads 
\begin{equation}
\mathbf{y}^{(j)}(t+1)=c\,\mathbf{P}^{(j)}\,\mathbf{y}^{(j)}(t)+(1-c)\,\mathbf{f}^{(j)},
\end{equation} 
where $\mathbf{y}^{(j)}$ and $\mathbf{f}^{(j)}$ are $(N-1)$-dimensional vectors, obtained by removing entry $j$. By writing the previous equation component by component, we obtain
\begin{equation}
y_i^{(j)}(t+1)=\frac{c}{k_i}\,\sum_{m\neq j} A_{im}\,y_{m}^{(j)}(t)+(1-c)\,f_i^{(j)}.
\label{fj_reduced}
\end{equation}
The last equation has a similar interpretation as equation~\eqref{fj}: each node (excluding node $j$) starts with an opinion $f^{(j)}_i$, and recursively updates it by considering its neighbors  opinions. Differently from equation~\eqref{fj}, node $j$  opinion does not contribute to the other nodes opinions. The stationary opinions $\mathbf{y}^{(j)}(c|\mathbf{f}^{(j)})$ satisfy the equation
\begin{equation}
\mathbf{y}^{(j)}(c|\mathbf{f}^{(j)})=c\,\mathbf{P}^{(j)}\,\mathbf{y}^{(j)}(c|\mathbf{f}^{(j)})+(1-c)\,\mathbf{f}^{(j)}.
\end{equation}
If $\mathbf{f}^{(j)}=\mathbf{p}^{(j)}c/(1-c)$ and $c=e^{-\lambda}$, where $\mathbf{p}^{(j)}$ and $\lambda$ are defined by the effective distance equation \eqref{effdist}, the solution to the previous equation is
\begin{equation}
\mathbf{y}^{(j)}(e^{-\lambda}|\mathbf{f}^{(j)})=(\mathbf{I}^{(j)}-e^{-\lambda}\,\mathbf{P}^{(j)})^{-1}\,e^{-\lambda} \mathbf{p}^{(j)}.
\end{equation}
Since the right-hand side of the equation is exactly equal to the partition function of effective distance \eqref{zeta},  in terms of the FJ social influence model the ViralRank centrality \eqref{viralrank} can be compactly written as 
\begin{equation}
v_i(\lambda) = -\frac{1}{N} \sum_{j} \ln \left({y}_i^{(j)}(e^{-\lambda}|\mathbf{f}^{(j)})\,{y}_j^{(i)}(e^{-\lambda}|\mathbf{f}^{(i)})\right),
\end{equation} 
where $\mathbf{f}^{(j)}=\mathbf{p}^{(j)}e^{-\lambda}/(1-e^{-\lambda})$ is the initial opinion of the FJ model with opinion  $j$ removed; $y_i^{(j)}$ is the final opinion of $i$ neglecting the contribution of node $j$, and analogously for $y_j^{(i)}$. 

The FJ opinion-formation process that leads to $y_i^{(j)}$ can be interpreted as follows: each
 node $i$ starts with an "opinion" proportional to $p_i^{(j)}= P_{ij}$ (with $i\neq j$) which represents the  probability of jumping from $i$ to $j$ in one time step. Each node iteratively updates its score by summing the probabilities $P_{im}$ of its neighbors, $j$ excluded, based on the FJ dynamics; the stationary state of this iterative process is $y_i^{(j)}$ which can be therefore interpreted as a (network-determined) effective transition probability $P_{ij}$. The ViralRank score $v_i$ of a given node $i$ therefore depends on all its effective transition probabilities $y_{i}^{(j)}$ and $y_j^{(i)}$.

\section{The relation between Google  PageRank and network effective distance}
\label{app:pr}

The PageRank score of a node is essentially a measure of how easy it is to reach a node with a random walk.
It is thus tempting to try to recover the PageRank vector of scores by modifying the effective distance in order to make it a measure of the reachability of a node for a diffusion process started from another node.

The PageRank vector is defined as the stationary density of a random-walk in discrete time on a graph and is described by the master equation~\cite{Gleich} 
\begin{align}
\pi_i(t+1) = c\sum_j \pi_j(t)\,P_{ji} +  (1-c)\,g_i,
\label{pr}
\end{align}
where $c \in (0,1)$ is the damping parameter, $\mathbf{g}$ is the \textit{preference} vector normalized to unity ($\sum_i g_i = 1$), and $\mathbf{P}$ is the row-stochastic transition probability matrix. The constant $(1-c)$ that multiplies the preference vector $\mathbf{g}$ gives the probability to jump to any random state, while the entry $g_i$ gives the conditional probability to teleport precisely to state $i$. 
The stationary solution of equation~\eqref{pr} reads  
\begin{align}
\mathbf{\pi} = \left(\mathbf{I}-c\mathbf{P}^T\right)^{-1}(1-c)\mathbf{g}.
\label{prstat}
\end{align}
In the most commonly used version of PageRank $g_i = 1/{N}$, $\forall i$, is the uniform distribution vector and $c=0.85$. Variants of this choice that consider a node dependent preference vector have been considered in  \cite{lambiotte2012ranking}.

To show the connection with effective distance, let us consider again the partition function \eqref{zeta}, which explicitly  reads
\begin{equation}
{Z}_{ij}(\mathbf{P},\lambda) = \sum_{k\ne j} \left( \mathbf{I}^{(j)}- e^{-\lambda} \mathbf{P}^{(j)}\right)^{-1}_{ik}  e^{-\lambda} p_k^{(j)}.
\label{modpart}
\end{equation}
Here $\mathbf{P}^{(j)}$ and $\mathbf{I}^{(j)}$ are the $(N-1)\times(N-1)$ submatrices of $(\mathbf{P})_{ij} = A_{ij}/\sum_k A_{ik}$ and of the identity $(\mathbf{I})_{ij} = \delta_{ij}$, respectively, obtained by excluding the $j$th row and $j$th column; $\mathbf{p}^{(j)}$ is the $j$th column of $\mathbf{P}$  after removing the $j$th row.
Let us now modify the previous equation and consider the alternative partition function
\begin{eqnarray}
\widetilde{Z}_{ij}(\mathbf{P},\lambda) = \sum_{k\ne j}\left( \mathbf{I}- e^{-\lambda} \mathbf{P}\right)^{-1}_{ik} e^{-\lambda} P_{kj}.
\label{ztilde}
\end{eqnarray}
Contrary to the partition function (\ref{modpart}), where only walks that terminate in $j$ are considered, in equation ~(\ref{ztilde}) also those walks that cross multiple times the target $j$ are considered.
By rearranging the sum for $\lambda >0$ we obtain
\begin{align}
\widetilde{Z}_{ij}(\mathbf{P}^{T},\lambda) 
&= 
\sum_{k\ne j}
\sum_{n=0}^\infty \left(e^{-\lambda} \mathbf{P}^T\right)^{n}_{ik} e^{-\lambda} P^T_{kj} 
\nonumber \\
&= \sum_{m\ne j}
\sum_{n=0}^\infty \sum_{k\ne j}e^{-\lambda}P_{jk} \left(e^{-\lambda} \mathbf{P}\right)^{n-1}_{km} e^{-\lambda}P_{mi} 
\nonumber \\
&= \sum_{m\ne j}
\sum_{n=0}^\infty  \left(e^{-\lambda} \mathbf{P}\right)^{n}_{jm} e^{-\lambda}P_{mi} 
\nonumber \\
&= 
\widetilde{Z}_{ji}(\mathbf{P},\lambda).
\label{}
\end{align} 
Then, by averaging the partition function \eqref{ztilde} over the source nodes $\{i\}$ we obtain the vector $\widetilde\pi_j = {N}^{-1}\sum_i \widetilde{Z}_{ij}(\mathbf{P})={N}^{-1}\sum_i\widetilde{Z}_{ji}(\mathbf{P}^T)$. Finally, if no self-loops are present we can include all terms in the sum \eqref{ztilde} so that $\widetilde\pi_j$ satisfies precisely the  PageRank equation \eqref{prstat} with dumping parameter $c=e^{-\lambda}$ and non-uniform preference vector \footnote{Note that in this form the preference vector is not yet normalized to unity.}  
\begin{eqnarray}
\widetilde{g}_k = \frac{1}{N}  \frac{e^{-\lambda} }{(1- e^{-\lambda})} \sum_i  P_{ik}.
\label{teleportation}
\end{eqnarray}

By contrast, ViralRank is built on the effective distance that depends logarithmically on a partition function that selects the walks that terminate once they hit the arrival node. We argue that these differences lead to the better ViralRank  performance for the toy network of Fig.~\ref{drawingb} and for the analyzed empirical networks for which we find that PageRank is even outperformed by degree in identifying influential spreaders \cite{SM}.

\section{Existing centrality measures}
\label{app:centralitymetrics}

\paragraph*{Degree centrality, $k$, and strength centrality, $s$.}
The degree centrality $k$ is arguably the simplest centrality measure, which is defined as the number of connections attached to each node. Given the adjacency matrix $\mathbf{A}$ -- $A_{ij} = 1$ if there is a
connection between nodes $i$ and $j$, zero otherwise -- the degree
centrality is the sum $k_i=\sum_j A_{ij}$. For weighted networks with weighted adjacency matrix $\mathbf{W}$ -- $W_{ij} \ge 0$ is the weight assigned to the
connection between nodes $i$ and $j$ --, the previous definition is naturally extended by the strength centrality $s_i=\sum_j W_{ij}$ \cite{dynamical}.

\paragraph*{$k$-core centrality, $k_c$.}
The $k$-core centrality \cite{kitsak2010identification} is obtained from the k-shell decomposition as the maximal connected subgraph composed of nodes that have at least $k$ neighbors within the set itself.  
Each node is endowed with an integer $k$-core index $k_c$ which equals the largest $k$ value of $k$-cores to which the node belongs.
This measure has been shown to outperform the degree centrality in the seminal work by Kitsak \textit{et al.} \cite{kitsak2010identification}.

\paragraph*{LocalRank, $LR$.}
LocalRank is a centrality that considers both the nearest and the next nearest neighbors to fourth order \cite{chen2012identifying}. It is defined as
\begin{equation}
[LR]_i = \sum_k A_{ik} \sum_m A_{km} \sum_n A_{mn} (1 + \sum_r A_{nr}).
\end{equation}
This metric has been shown to be competitive in the influential spreaders identification by L\"u \textit{et al.} \cite{lu2016vital}.

\paragraph*{Non-backtracking centrality, $NBC$.}

The non-backtracking centrality \cite{martin2014localization} is introduced to overcome the limitation of the eigenvector centrality by considering the
Hashimoto or non-backtracking matrix  \cite{hashimoto1989zeta,krzakala2013spectral}. Given an abstract undirected network with $E$ edges,
we construct a directed version of it  with $2E$ edges, where each original edge has been replaced by two directed ones pointing in opposite directions.  The non-backtracking matrix $\mathbf{B}$ is the $2E \times 2E$  matrix, where each element corresponds to a pair of directed edges, defined as 
\begin{equation}
B_{i \rightarrow j, k\rightarrow l } = \delta_{jk}(1-\delta_{il}).
\label{nbmatrix}
\end{equation}
Thus, the only non-zero elements of $\mathbf{B}$ are those defining non-backtracking paths of lengths two, from $i$ to $l$ via $j$, with $j=k$ and $l\ne i$.

The non-backtracking centrality is defined as 
\begin{equation}
[NBC]_i = \sum_j A_{ij} v_{i\rightarrow j},
\end{equation}
where $v_{i\rightarrow j}$ is the eigenvector corresponding to the largest eigenvalue of the non-backtracking matrix \eqref{nbmatrix}.
For the Perron-Frobenius theorem the largest eigenvalue of $\mathbf{B}$ is always real and positive and so are the components of the corresponding eigenvector.
A much faster calculation of the non-backtracking centrality can be carried out via the Ihara-Bass
determinant as the first $N$ elements of the leading left eigenvector of the $2N \times 2N$ matrix \cite{krzakala2013spectral}
\begin{equation}
\mathbf{B'}=
\begin{pmatrix}
\mathbf{0} & \mathbf{K}-\mathbf{I}  \\
-\mathbf{I} & \mathbf{A} 
\end{pmatrix}
\end{equation}
where $K_{ij} = \delta_{ij} k_i$ is the diagonal matrix with the degrees $k_i$ as entries and $(\mathbf{I})_{ij} = \delta_{ij}$ is the identity matrix.

Radicchi and Castellano  showed that the non-backtracking centrality is the most competitive metric to identify the influential spreaders for spreading processes at criticality \cite{radicchi2016leveraging}.

\paragraph*{Random-walk accessibility, $RWA$.}

The (generalized) random-walk accessibility  \cite{travenccolo2008accessibility,de2014role} is a measure that quantifies the diversity of access of individual nodes via random
walks. The accessibility is defined by the
exponential of the Shannon entropy 
\begin{equation}
[RWA]_i = \exp \left( -\sum_j M_{ij} \ln M_{ij} \right),
\end{equation}
where $\mathbf{M} = \exp(\mathbf{P})$ takes into account walks of growing length on the network.
Thus, by construction the accessibility penalizes longer walks. 

The metric has been shown to be competitive for the influential spreaders identification in geographically embedded networks \cite{de2014role}.

\section{Details on the empirical datasets}
\label{app:datasets}

The empirical datasets used for the \emph{contact-network} dynamics are: (1) [\textit{Terrorists}] The terrorist network~\cite{krebs2002mapping} which includes the terrorists (nodes) who belonged to the terroristic cell components centered around the $19$ dead hijackers involved in the attacks at the World Trade Center on September 11th, 2001. Each link identifies a social or economic interaction between two terrorists. (2) [\textit{Email}] The email contact network~\cite{Michalski2011} where the nodes represent employees of a mid-sized manufacturing company. Two employees are connected by a link if they exchanged at least one email in the year 2010. (3) [\textit{Jazz}] In the jazz collaboration network~\cite{gleiser2003community} the nodes represent jazz musicians and the links represent their recorded collaborations between 1912 and 1940.; (4) [\textit{NetSci}] This is the largest connected component of the network scientists  co-authorship network~\cite{newman2006finding} where the nodes are scientists working on network theory and experiments. Two scientists are linked if they co-authored at least one paper in the years prior to  2007. (5) [\textit{Protein}] The network of interactions between the proteins contained in yeast~\cite{konect:coulomb2005}; each node represents a protein, and an edge represents a metabolic interaction between two proteins. (6) [\textit{Facebook}] A Facebook friendship network~\cite{leskovec2012learning} where the nodes represent Facebook users, and the links represent their friendship relations collected from survey participants.

\section*{Acknowledgements}
The authors thank Linyuan L\"u for her detailed feedback on the manuscript and for several discussions on the topic and Giulio Iannelli for technical assistance in producing Fig. 1.

This work has partially been funded by the DFG / FAPESP, within the scope of the IRTG 1740 / TRP 2015/50122-0.
M.S.M. acknowledges financial support from the
Science Strength Promotion Programme of the UESTC, from the Universit\"at Z\"urich through the URPP Social Networks, and from
the Swiss National Science Foundation Grant No. 200020-156188.

\bibliography{sample}

\end{document}